\documentclass[10pt, conference]{IEEEtran}
\usepackage{url}
\usepackage{multirow}
\usepackage{mdframed}
\usepackage{lipsum}
\usepackage{setspace}
\usepackage{graphicx}
\usepackage{graphics}
\usepackage{latexsym}
\usepackage{showidx}
\usepackage{latexsym}
\usepackage{amssymb}
\usepackage{amsfonts}
\usepackage{amsmath}
\usepackage{amssymb}
\usepackage{amsfonts}
\usepackage{amsmath}
\IEEEoverridecommandlockouts

\newtheorem{remark}{Remark}

\usepackage[tight,footnotesize]{subfigure}
\usepackage{stfloats}
\begin{document}

\title{On Complex LLL Algorithm for Integer Forcing Linear Receivers}
\author{\IEEEauthorblockN{A.~Sakzad, J.~Harshan, and E.~Viterbo,
\thanks{This work was performed at the Monash Software Defined Telecommunications Lab and was supported by the Monash Professional Fellowship and the Australian Research Council under
Discovery grants ARC DP 130100103.}}
\IEEEauthorblockA{Department of ECSE, Monash University, Australia}
{\tt $\{$amin.sakzad, harshan.jagadeesh, and emanuele.viterbo$\}$@monash.edu}}
%\and

\maketitle
%%------------------------------------------------------------------------------------
%%------------------------------------------------------------------------------------
\begin{abstract}
Integer-forcing (IF) linear receiver has been recently introduced for multiple-input multiple-output (MIMO) fading channels. The receiver has to compute an integer linear combination of the symbols as a part of the decoding process. In particular, the integer coefficients have to be chosen based on the channel realizations, and the choice of such coefficients is known to determine the receiver performance. The original known solution of finding these integers was based on exhaustive search. A practical algorithm based on Hermite-Korkine-Zolotareff (HKZ) and Minkowski lattice reduction algorithms was also proposed recently. In this paper, we propose a low-complexity method based on complex LLL algorithm to obtain the integer coefficients for the IF receiver. For the $2 \times 2$ MIMO channel, we study the effectiveness of the proposed method in terms of the ergodic rate. We also compare the bit error rate (BER) of our approach with that of other linear receivers, and show that the suggested algorithm outperforms the minimum mean square estimator (MMSE) and zero-forcing (ZF) linear receivers, but trades-off error performance for complexity in comparison with the IF receiver based on exhaustive search or on HKZ and Minkowski lattice reduction algorithms.
\end{abstract}
%%%%%%%%%%%%%%%%%%%%%%%%%%%%%%%%%%%%%%%%%%%%%%%%%%%%%%%%%%%%%%%%%%%%%%%%%%%%%%%%%%%%%%
%%%%%%%%%%%%%%%%%%%%%%%%%%%%%%%%%%%%%%%%%%%%%%%%%%%%%%%%%%%%%%%%%%%%%%%%%%%%%%%%%%%%%%
\section{Introduction}
In multipath fading channels, using multiple antennas at the transceivers is known to provide large capacity gains. Such a capacity gain comes at the cost of high decoding complexity at the receiver. It is known that a high-complexity joint ML decoder can be employed at the receiver to reliably recover the information. On the other hand, the \emph{linear receivers} such as the ZF and the MMSE receiver~\cite{Kumar09} reduce the decoding complexity trading-off error performance.

The integer forcing (IF) linear receiver has been recently proposed~\cite{zhan12}. This new architecture obtains high rates in MIMO fading channels. In this approach, the transmitter employs a layered structure with identical lattice codes for each layer. Then each receive antenna is allowed to find an integer linear combination of transmitted codewords. The decoded point will be another lattice point because any integer linear combination of lattice points is another lattice point. This idea has been brought to MIMO channels from the compute-and-forward protocol for physical layer network coding~\cite{Nazer11, Narayanan10}.
%\footnote{A lattice $\Lambda$ with basis $\{{\bf g}_1,\ldots,{\bf g}_n\}\subseteq\mathbb{C}^n$, is the set of all points
%$\{{\bf x}={\bf u}{\bf G}| {\bf u}\in \mathbb{Z}[i]^n\}$ where the generator matrix ${\bf G}$ of $\Lambda$ is a matrix formed by placing ${\bf g}_m$'s as its rows.
%The Gram matrix of $\Lambda$ is ${\bf M}={\bf G}{\bf G}^h$.}

In the MIMO IF architecture, a filtering matrix $\textbf{B}$ and a non-singular integer matrix ${\bf A}$ are needed such that ${\bf B}{\bf H} \approx {\bf A}$ with minimum quantization error at high signal-to-noise ratio ($\mbox{\small SNR}$) values. The exhaustive search solution to the problem of finding ${\bf A}, {\bf B}$ is addressed in~\cite{zhan12}. It is prohibitively complex already for $2\times 2$ real MIMO and becomes untractable for $2\times2$ complex MIMO and beyond. A smart practical method of finding ${\bf A}$ based on HKZ and Minkowski lattice reduction algorithms has been proposed recently~\cite{SHV:submitted}. This provides full receive diversity with much lower complexity in comparison to exhaustive search. The major differences between integer-forcing linear receivers and lattice reduction aided MIMO detectors~\cite{Taherzadeh07-1,Taherzadeh07-2,matz11} are also presented in~\cite{SHV:submitted}.

In this paper, we propose a low-complexity method for choosing the above matrices. In \cite{SHV:submitted,zhan12}, a $2n$-layered scheme is considered with real lattice codebook for each layer. Unlike the model there, we work on complex entries and we lift that set-up to complex case. The proposed method is a combination of three low-complexity methods which are based on complex lattice reduction (CLLL)~\cite{CLLL09} technique for a lattice, and singular value decomposition (SVD) of matrices. For the $2 \times 2$ MIMO channel, we compare the performance (in terms of ergodic rate and uncoded probability of error) of the proposed low-complexity solution with the known linear receivers and show that the proposed solution ({\em i}) provides a lower bound on the ergodic rate of the IF receiver, ({\em ii}) outperforms the ZF and MMSE receivers in probability of error, and ({\em iii}) trades off error performance for computational complexity in comparison with exhaustive search and other lattice reduction methods including HKZ and Minkowski algorithms.

The rest of the paper is organized as follows. In Section~\ref{sec:back}, we give a brief background on lattices. We present the problem statement along with the signal model in Section~\ref{sec:model}. In Section~\ref{sec:methods}, we study the solution to the IF receiver via two CLLL algorithms. A complexity comparison for different known approaches is also given in this section. In Section~\ref{sec:simulations}, we show some simulation results on the performance of IF receiver in ergodic MIMO setting. Finally, we present concluding remarks in Section~\ref{sec:conclusion}.

{\em Notations}.  Row vectors are presented by boldface letters, and matrices are denoted by capital boldface letters. Let ${\bf v}$ be a vector, ${\bf v}^T$ denotes transposition, and ${\bf v}^h$ denotes the Hermitian transposition.
%A matrix formed by attaching the row vectors $ \{ \textbf{x}_{m} ~|~1\leq m\leq n \}$ is denoted by $\left[\textbf{x}_{1}^T,\ldots,\textbf{x}_{n}^T\right]^T$.
We show the $n\times n$ identity and zero matrix as ${\bf I}_n$ and ${\bf 0}_n$ respectively. For a matrix $\textbf{X}$, the element in the $k$-th row and $m$-th column of $\textbf{X}$ will be denoted by $\textbf{X}_{k, m}$.
The sets $\mathbb{C}$, and $\mathbb{Z}[i]$ denote the set of all complex numbers, and the Gaussian integer ring, respectively, where $i = \sqrt{-1}$.  If $z\in\mathbb{C}$, then $\Re{(z)}$ is the real part and $\Im{(z)}$ is the imaginary part of $z$. Let $|\cdot|$ denote the modulus of a complex number. The $\| \cdot \|$ operation denotes the norm square of a vector. For a complex number $z$, the closest Gaussian integer to $z$ is denoted by $\lfloor z \rceil$, which we refer as the quantization of $z$. The notation $\lfloor {\bf v} \rceil$ is the component-wise quantized version of the vector ${\bf v}$. The Hermitian product of ${\bf a}$ and ${\bf b}$ is defined by $\langle{\bf a}, {\bf b}\rangle \triangleq {\bf b}^{h} {\bf a}$. Finally, the set of orthogonal vectors generated by the Gram-Schmidt orthogonalization procedure are denoted by $\{\mbox{GS}({\bf d}_1),\ldots,\mbox{GS}({\bf d}_n)\}$.

%%%%%%%%%%%%%%%%%%%%%%%%%%%%%%%%%%%%%%%%%%%%%%%%%%%%%%%%%%%%%%%%%%%%%%%%%%%%%%%%%%%%%%
%%%%%%%%%%%%%%%%%%%%%%%%%%%%%%%%%%%%%%%%%%%%%%%%%%%%%%%%%%%%%%%%%%%%%%%%%%%%%%%%%%%%%%
\section{Background on Lattices and CLLL Algorithm}\label{sec:back}
%%%%%%%%%%%%%%%%%%%%%%%%%%%%%%%%%%%%%%%%%%%%%%%%%%%%%%%%%%%%%%%%%%%%%%%%%%%%%%%%%%%%%%
%%%%%%%%%%%%%%%%%%%%%%%%%%%%%%%%%%%%%%%%%%%%%%%%%%%%%%%%%%%%%%%%%%%%%%%%%%%%%%%%%%%%%%
A lattice $\Lambda$ with basis
$\{{\bf g}_1,{\bf g}_2,\ldots,{\bf g}_n\}$, where ${\bf g}_k \in \mathbb{C}^n$,
is the set of all points $\{{\bf x}={\bf u}{\bf G}| {\bf u}\in {\mathbb{Z}[i]}^n\}$.
A generator matrix for $\Lambda$ is an $n\times n$ complex matrix
${\bf G}=\left[{\bf g}_1^T,\ldots,{{\bf g}_n}^T\right]^T$.
The Gram matrix of $\Lambda$ is ${\bf M}={\bf G}{\bf G}^h$.
The $m$--th successive minima of $\Lambda$, denoted by $\lambda_m$, is the radius of the
smallest possible closed ball around origin containing $m$ or more linearly independent
lattice points.

In complex lattice reduction, we let ${\bf G}'={\bf U}{\bf G}$,
where ${\bf U}$ is an unimodular matrix. Let us define
\[
\mu_{\ell,j}=\frac{\langle{\bf g}_{\ell},\mbox{GS}\left({\bf g}'_j\right)\rangle}{\|\mbox{GS}\left({\bf g}'_j\right)\|^2}
\]
where $1\leq \ell,j \leq n$.
A generator matrix
${\bf G}'$ is said to be CLLL-reduced if the following two conditions are satisfied~\cite{CLLL09}:
\begin{enumerate}
  \item for $1\leq j<\ell\leq n$ \[ |{\Re}(\mu_{\ell,j})|\leq 1/2, ~~~~~ |{\Im}(\mu_{\ell,j})|\leq 1/2,\]
  \item for $1< m\leq n$, \[\|\mbox{GS}\left({\bf g}'_m\right)\|^2\geq \left(\delta-|\mu_{m,m-1}|^2\right)\|\mbox{GS}\left({\bf g}'_{m-1}\right)\|^2 \]
where $\delta \in (1/2, 1]$ is a factor selected
to achieve a good quality-complexity tradeoff.
\end{enumerate}
An algorithm is provided in~\cite{CLLL09} to evaluate a CLLL-reduced basis matrix $\bf{G}'$ of a lattice $\Lambda$ with a generator matrix ${\bf G}$. The input of this algorithm is the matrix $\bf{G}$ and a factor $\delta$, and the outputs of the algorithm are the unimodular matrix ${\bf U}$ and the CLLL-reduced basis matrix ${\bf G}'$ such that ${\bf G}'={\bf U}{\bf G}$.
%It was shown in \cite{CLLL09} that the rows of ${\bf G}'$, denoted by ${\bf g}'_m$, satisfies
%\begin{equation}~\label{successive}
%\|{\bf g}'_m\|^2\leq \alpha^{n-1}\lambda_m,~~~~~~1\leq m\leq n.
%\end{equation}
%where $\alpha=1/(\delta-1/2)$.
%%%%%%%%%%%%%%%%%%%%%%%%%%%%%%%%%%%%%%%%%%%%%%%%%%%%%%%%%%%%%%%%%%%%%%%%%%%%%%%%%%%%%%
%%%%%%%%%%%%%%%%%%%%%%%%%%%%%%%%%%%%%%%%%%%%%%%%%%%%%%%%%%%%%%%%%%%%%%%%%%%%%%%%%%%%%%
\section{Signal Model and Problem Statement}~\label{sec:model}
%%%%%%%%%%%%%%%%%%%%%%%%%%%%%%%%%%%%%%%%%%%%%%%%%%%%%%%%%%%%%%%%%%%%%%%%%%%%%%%%%%%%%%
%%%%%%%%%%%%%%%%%%%%%%%%%%%%%%%%%%%%%%%%%%%%%%%%%%%%%%%%%%%%%%%%%%%%%%%%%%%%%%%%%%%%%%
A flat-fading MIMO channel with $n$ transmit antennas and $n$ receive antennas is considered. The channel matrix ${\bf H}$ is in $\mathbb{C}^{n\times n}$, where the entries of ${\bf H}$ are i.i.d. as $\mathcal{CN}(0, 1)$  this channel coefficient remains fixed for a given interval (of at least $N$ complex channel uses) and take an independent realization in the next interval. We use a $n$-layer transmission scheme where the information transmitted across different antennas are statistically independent. For $1 \leq m \leq n$, the $m$-th layer is equipped with an encoder $\mathcal{E}_m:\mathcal{R}^{k}\rightarrow\mathbb{C}^{N}$. This encoder maps a vector message ${\bf m}\in \mathcal{R}^{k}$, where $\mathcal{R}$ is a ring, into a lattice codeword ${\bf x}_m \in \Lambda \subset \mathbb{C}^{N}$. If ${\bf X}$ denotes the matrix of transmitted vectors, the received signal ${\bf Y}$ is given by ${\bf Y} = \sqrt{P}{\bf H}{\bf X}+{\bf Z}$, where $P=\frac{\mbox{\small SNR}}{n}$ and ${\mbox\small SNR}$ denotes the average signal-to-noise ratio at each receive antenna. The entries of ${\bf Z}$ are i.i.d. and distributed as $\mathcal{CN}(0, 1)$. We also assume that the channel state information is only available at the receiver. The goal of IF linear receiver is to approximate ${\bf H}$ with a non-singular integer matrix ${\bf A}$. Since we suppose the information symbols to be in the ring $\mathcal{R}$, we look for an invertible matrix ${\bf A}$ over the ring $\mathcal{R}$. Thus, we have $${\bf Y}'={\bf B}{\bf Y}=\sqrt{P}{\bf A}{\bf X}+\sqrt{P}({\bf B}{\bf H}-{\bf A}){\bf X}+{\bf B}{\bf Z}.$$
As $\sqrt{P}{\bf A}{\bf X}$ is the useful signal component, the effective noise is $\sqrt{P}({\bf B}{\bf H}-{\bf A}){\bf X}+{\bf B}{\bf Z}$. In particular, the power of the $m$-th row of effective noise is $\|{\bf b}_m\|^2 + P\|{\bf b}_m{\bf H}-{\bf a}_m\|^2$. Hence, we define
\begin{equation}~\label{quntizederrplusnoise}
g({\bf a}_m,{\bf b}_m)\triangleq \|{\bf b}_m\|^2 + P\|{\bf b}_m{\bf H}-{\bf a}_m\|^2,
\end{equation}
where ${\bf a}_m$ and ${\bf b}_m$ denote the $m$-th row of ${\bf A}$ and $\bf{ B}$, respectively. Note that in order to increase the effective signal to noise ratio for each layer, the term $g({\bf a}_m,{\bf b}_m)$ has to be minimized for each $m$ by appropriately selecting the matrices ${\bf A}$ and ${\bf B}$. We formally put forth the problem statement below (see~\cite{zhan12}):
\vspace{-.4cm}
\begin{mdframed}
Given ${\bf H}$ and $P$, the problem is to find the matrices ${\bf B} \in \mathbb{C}^{n \times n}$ and ${\bf A} \in \mathbb{Z}[i]^{n \times n}$ such that
\begin{itemize}
\item the $\max_{1\leq m\leq n}g({\bf a}_m,{\bf b}_m)$ is minimized, and
\item the matrix ${\bf A}$  is invertible over the ring $\mathcal{R}$.
\end{itemize}
\end{mdframed}
\vspace{.1cm}
In order to have ZF receiver, we put ${\bf B}={\bf H}^{-1}$ and ${\bf A}={\bf I}_n$. If we let ${\bf B}={\bf H}^h{\bf S}^{-1}$ where
\begin{equation}
\label{s_matrix}
{\bf S}=P^{-1}{\bf I}_n+{\bf H}{\bf H}^h,
\end{equation}
and ${\bf A}={\bf I}_n$, then we get the linear MMSE receiver.

In \cite{zhan12}, the authors have proposed a method to obtain ${\bf A}$ and ${\bf B}$, which reduces $g({\bf a}_m,{\bf b}_m)$ for each $m$. We now recall the approach presented in \cite{zhan12}. First, conditioned on a fixed ${\bf a}_m = \bf{a}$, the term $g({\bf a},{\bf b}_m)$ is minimized over all possible values of ${\bf b}_m$. As a result, the optimum value of ${\bf b}_m$ can be obtained as
\begin{equation}
\label{b_vector}
{\bf b}_m={\bf a}{\bf H}^h{\bf S}^{-1}.
\end{equation}
Then, after replacing ${\bf b}_m$ of \eqref{b_vector} in $g({\bf a},{\bf b}_m)$, one has to minimize $g({\bf a},{\bf a}{\bf H}^h{\bf S}^{-1})$ over all possible values of ${\bf a}$ to obtain ${\bf a}_{m}$ as ${\bf a}_{m} = \arg \min_{{\bf a}} g({\bf a},{\bf a}{\bf H}^h{\bf S}^{-1})$. The last expression can be written as
\begin{equation}
\label{opt_problem}
{\bf a}_{m} = \arg \min_{{\bf a}} ~{\bf a}{\bf V}{\bf D}{\bf V}^h{\bf a}^h,
\end{equation}
where ${\bf V}$ is the matrix composed of the eigenvectors of ${\bf H}{\bf H}^{h}$ and ${\bf D}$ is a diagonal matrix with $m$-th entry ${\bf D}_{m,m}=\left(P\rho_m^2+1\right)^{-1}$, where $\rho_m$ is the $m$-th singular value of ${\bf H}$. With this, we have to obtain $n$ vectors $\{ {\bf a}_{m}\}$ which result in the first $n$ smaller values of ${\bf a}{\bf V}{\bf D}{\bf V}^h{\bf a}^h$ along with the non-singular property on ${\bf A}$. In order to get ${\bf a}_m$, $1\leq m\leq n$, the authors of ~\cite{zhan12} have suggested an exhaustive search for each component of ${\bf a}_m$
within a sphere of squared radius
\begin{equation}\label{radius}
1+P\rho_{\max}^2,
\end{equation}
where $\rho_{\max}=\max_{m}\rho_m$. It has also been pointed out in~\cite{zhan12} that this search can be accelerated by means of a sphere decoder on the lattice with Gram matrix ${\bf M}' = {\bf V}{\bf D}{\bf V}^h$. For a fixed $P$, the complexity of this approach is of order $O\left(P^n\right)$. It is also shown in~\cite{zhan12} that the exhaustive search approach provides a diversity order of $n$ and a multiplexing gain of $n$. The outage rate and probability of this scheme are studied in~\cite{zhan12}. At this stage, we note that the exhaustive computation of ${\bf a}_m$ has high computational complexity, especially for large values of $P$ and $n$, and hence the approach in~\cite{zhan12} is not practical even for the $2\times 2$ complex case.

The best possible ${\bf a}_m$'s for this problem are the set of all successive minimas of ${\bf V}{\bf D}^{\frac{1}{2}}$, which can be approximately computed by either the Minkowski or HKZ lattice reduction algorithms~\cite{korkine1873,Minkowski1891,zhang12}. In~\cite{SHV:submitted}, using HKZ and Minkowski reduction algorithms, we obtain the matrices ${\bf A}$ and ${\bf B}$ and show that these practical algorithms achieve full receive diversity in terms of error performance with reasonable complexity. Since the CLLL algorithm has lower complexity than HKZ and Minkowski reduction algorithms, it could be employed to get matrix ${\bf A}$ as well. Hence, in the rest of the paper we focus on using CLLL algorithm to find the best possible ${\bf A}$ for our problem.
%% because the radius of the mentioned sphere
%%and the number of integer components to be obtained are increasing by $P$ and $n$.
%Although, joint minimization over ${\bf A}$ and ${\bf B}$ has not been addressed.
%Finally, the rows of the matrix ${\bf B}$ proposed in \cite{zhan12} is of the form
%
%Unlike the model in \cite{zhan12},
%we work on complex matrices, and hence the goal is to obtain a matrix $\textbf{A}$
%over Gaussian integers $\mathbb{Z}[i]$ instead of a matrix over integers.
%The method proposed in \cite{zhan12} is the only known method to obtain ${\bf A}$ and ${\bf B}$.
%Note that, since the channel $\textbf{H}$ changes for every quasi-static interval,
%${\bf A}$ and ${\bf B}$ matrices have to be computed at the receiver in the beginning
%of every quasi-static interval. Next we present the ergodic rate of
%the IF linear receivers and discuss the impact of the choice of ${\bf A}$ and ${\bf B}$ on the ergodic rate.
%%%%%%%%%%%%%%%%%%%%%%%%%%%%%%%%%%%%%%%%%%%%%%%%%%%%%%%%%%%%%%%%%%%%%%%%%%%%%%%%%%%%%%
%%%%%%%%%%%%%%%%%%%%%%%%%%%%%%%%%%%%%%%%%%%%%%%%%%%%%%%%%%%%%%%%%%%%%%%%%%%%%%%%%%%%%%
%%%%%%%%%%%%%%%%%%%%%%%%%%%%%%%%%%%%%%%%%%%%%%%%%%%%%%%%%%%%%%%%%%%%%%%%%%%%%%%%%%%%%%
%%%%%%%%%%%%%%%%%%%%%%%%%%%%%%%%%%%%%%%%%%%%%%%%%%%%%%%%%%%%%%%%%%%%%%%%%%%%%%%%%%%%%%
%%%%%%%%%%%%%%%%%%%%%%%%%%%%%%%%%%%%%%%%%%%%%%%%%%%%%%%%%%%%%%%%%%%%%%%%%%%%%%%%%%%%%%
%%%%%%%%%%%%%%%%%%%%%%%%%%%%%%%%%%%%%%%%%%%%%%%%%%%%%%%%%%%%%%%%%%%%%%%%%%%%%%%%%%%%%%
\section{Low-complexity IF Receivers}~\label{sec:methods}
In this section, we propose three low-complexity methods to obtain some candidates for the rows of ${\bf A}$. The first two are based on CLLL algorithm and the last one is based on SVD decomposition. Then, we propose a selective combining technique to choose the rows of ${\bf A}$ from the candidate rows. Once we obtain ${\bf A}$, we obtain ${\bf B}$ as ${\bf B}={\bf A}{\bf H}^h{\bf S}^{-1}$, where ${\bf S}$ is given in \eqref{s_matrix}. Henceforth, we only address systematic methods to obtain ${\bf A}$.
%%%%%%%%%%%%%%%%%%%%%%%%%%%%%%%%%%%%%%%%%%%%%%%%%%%%%%%%%%%%%%%%%%%%%%%%%%%%%%%%%%%%%%
%%%%%%%%%%%%%%%%%%%%%%%%%%%%%%%%%%%%%%%%%%%%%%%%%%%%%%%%%%%%%%%%%%%%%%%%%%%%%%%%%%%%%%
\subsection{Algorithm 1 via CLLL}
It is pointed out in \cite{zhan12} that the minimization problem in \eqref{opt_problem} is the shortest vector problem for a lattice with Gram matrix ${\bf M}'= {\bf V}{\bf D}{\bf V}^h$. Since ${\bf M}'$ is a positive definite matrix, we can write ${\bf M}' = {\bf L}{\bf L}^h$ for some ${\bf L} \in \mathbb{C}^{n \times n}$ by using Choelsky decomposition. With this, the rows of ${\bf L}={\bf V}{\bf D}^{\frac{1}{2}}$ generate a lattice, say $\Lambda'$. Based on \eqref{opt_problem}, a set of possible choices for $\{{\bf a}_1, \ldots, {\bf a}_n \}$ is the set of complex integer vectors whose corresponding lattice points in $\Lambda'$ have lengths at most equal to the $n$-th successive minima of $\Lambda'$. However, finding these vectors is again computationally complex.
\vspace{-.3cm}
\begin{mdframed}
{\bf Input:} ${\bf H} \in \mathbb{C}^{n\times n}$, and $P$.\\
{\bf Output:} A set of $n$ candidates for the rows of ${\bf A}$.
\begin{enumerate}
\item{} Form ${\bf L}={\bf V}{\bf D}^{\frac{1}{2}}$ of a lattice $\Lambda'$.
\item{} Reduce ${\bf L}$ to ${\bf L}^{\mbox\small clll}$ by CLLL algorithm.
\item{} The $n$ rows of ${\bf L}^{\mbox\small clll}{\bf L}^{-1}$ provides $n$ values of ${\bf a}_m$.
\end{enumerate}
\end{mdframed}
\vspace{.1cm}
We now use the CLLL algorithm~\cite{CLLL09} to obtain the complex integer vectors. In particular, we use the CLLL algorithm to reduce the basis set in ${\bf L}$ to obtain a new basis set represented by the rows of ${\bf L}^{\mbox\small clll}$. For each $1 \leq m \leq n$, it is known that
the length of the $m$-th row vector in ${\bf L}^{\mbox\small clll}$ is upper bounded by a scaled version of the $m$-th successive minima of $\Lambda'$~\cite{CLLL09}. Hence, the rows of ${\bf L}^{\mbox\small clll}{\bf L}^{-1}$ can be used to obtain $n$ possible choices for the desired matrix ${\bf A}$. %We summarize the above algorithm:
We note that the structure of the above algorithm is exactly the same as the one presented in~\cite{SHV:submitted} with only one difference: we used HKZ and Minkowski lattice reduction algorithms instead of CLLL algorithm.
\subsection{Algorithm $2$ via CLLL}
Given ${\bf H}$ and $P$, let us define a $2n-$dimensional complex lattice $\Lambda$
generated by
\begin{equation}~\label{G}
{\small {\bf G}=\left[\begin{array}{c|c}
P^{-1/2}{\bf I}_n &-{\bf H}\\
\hline
{\bf 0}& {\bf I}_n
\end{array}\right]  \in \mathbb{C}^{2n \times 2n}.}
\end{equation}
The Gram matrix of this lattice is ${\bf M}={\bf G}{\bf G}^h$. The Schur's component of ${\bf M}$ denoted by $({\bf M}|{\bf S})$ is given by $({\bf M}|{\bf S})={\bf I}_n-{\bf H}^h{\bf S}^{-1}{\bf H}$, while the Schur's decomposition of ${\bf M}$ can be derived as,
$${\small{\bf M}=\left[\begin{array}{c|c}
{\bf I}_n &{\bf 0}_{n}\\
\hline
-{\bf H}^h{\bf S}^{-1}& {\bf I}_n
\end{array}\right]
\left[\begin{array}{c|c}
{\bf S} &{\bf 0}_{n}\\
\hline
{\bf 0}_{n} & ({\bf M}|{\bf S})
\end{array}\right]
\left[\begin{array}{c|c}
{\bf I}_n &-{\bf S}^{-1}{\bf H}\\
\hline
{\bf 0}_{n} & {\bf I}_n
\end{array}\right]}.$$
Replacing ${\bf H}$ by its SVD representation,
one can easily observe that
${\bf M}'=({\bf M}|{\bf S})$.
Consider ${\bf x} \in \Lambda$ where ${\bf x} = {\bf u}{\bf G}$
and ${\bf u} \in \mathbb{Z}[i]^{2n}$ and
${\bf u} = \left[\textbf{b}|\textbf{a}\right]$
forming by adjoining row vector ${\bf a}$ after ${\bf b}$
for some $\textbf{a}, \textbf{b} \in \mathbb{Z}[i]^n$.
%the squared norm of ${\bf w}$ denoted by $f({\bf c}, {\bf d})$ is given by
For ${\bf u} = \left[{\bf d}|{\bf c}\right]$ with ${\bf d}, {\bf c} \in \mathbb{Z}[i]^n$, we define
$f({\bf c}, {\bf d}) \triangleq {\bf u}{\bf M}{\bf u}^h$.
Further, we expand the term $f({\bf c}, {\bf d})$ to obtain $f({\bf c}, {\bf d}) = P^{-1}\|{\bf d}\|^2+\|{\bf d}{\bf H}-{\bf c}\|^2$.
Note that $f({\bf c}, {\bf d}) = P^{-1}g({\bf c}, {\bf d})$. Therefore, the solutions to the minimization of $f({\bf c}, {\bf d})$ are also the solutions to the minimization of $g({\bf c}, {\bf d})$. Towards minimizing $f({\bf c}, {\bf d})$, one can immediately recognize that solving minimization of $f({\bf c}, {\bf d})$ is nothing but finding the shortest vector of $\Lambda$.
%Instead of exhaustively searching for the short vectors of $\Lambda$, we apply the CLLL algorithm to obtain the short vectors.
%we choose the optimal projection matrix ${\bf B}$ for a given matrix ${\bf A}$
%%as ${\bf B}={\bf A}{\bf H}^h{\bf S}^{-1}$.
%If we apply the relation \eqref{b_vector} in $f({\bf a}_m, {\bf b}_m)$,then the term in (\ref{firsttermerror}) will vanish. As a result, $f({\bf a}_m, {\bf b}_m)$ becomes ${\bf a}_m{\bf L}{\bf a}_m^h$. With this, we should find ${\bf a}_m$ such that ${\bf a}_m{\bf L}{\bf a}_m^h$ in \eqref{secondtermerror} is minimized. We now present a procedure for finding the matrix ${\bf A}$ by applying the CLLL algorithm on ${\bf G}$.\\
%%%%%%%%%%%%%%%%%%%%%%%%%%%%%%%%%%%%%%%%%%%%%%%%%%%%%%%%%%%%%%%%%%%%%%%%%%%%%%%%%%%%%%
%%%%%%%%%%%%%%%%%%%%%%%%%%%%%%%%%%%%%%%%%%%%%%%%%%%%%%%%%%%%%%%%%%%%%%%%%%%%%%%%%%%%%%
%%%%%%%%%%%%%%%%%%%%%%%%%%%%%%%%%%%%%%%%%%%%%%%%%%%%%%%%%%%%%%%%%%%%%%%%%%%%%%%%%%%%%%
%%%%%%%%%%%%%%%%%%%%%%%%%%%%%%%%%%%%%%%%%%%%%%%%%%%%%%%%%%%%%%%%%%%%%%%%%%%%%%%%%%%%%%
For the matrix ${\bf G}$ in \eqref{G}, let ${\bf G}^{\mbox \small clll}$ denote the $2n$--dimensional CLLL-reduced generator matrix of $\Lambda$. Using the short vectors in ${\bf G}^{\mbox\small clll}$, we obtain the complex integer matrix ${\bf U}$ with $2n$ row vectors ${\bf u}_{m} = \left[{\bf d}_{m}|{\bf c}_{m}\right]$ such that ${\bf U} = {\bf G}^{\mbox\small clll}{\bf G}^{-1}$. With this, we have obtained complex integer vectors ${\bf d}_{m}, {\bf c}_{m}$ which results in smaller values for $g({\bf d}, {\bf c})$. Hence, the vectors ${\bf d}_{m}, {\bf c}_{m}$ can be readily used for the IF architecture as ${\bf b}_{m} = {\bf d}_{m}$, and ${\bf a}_{m} = {\bf c}_{m}$. However, we do not use ${\bf b}_{m}$ as it is a complex integer vector. In such a case, ${\bf b}_{m} {\bf H} \neq {\bf a}_{m} $ even for large $\mbox{\small SNR}$ values which in turn results in an error floor in the probability of error. Instead, we only use ${\bf a}_{m}$, and subsequently obtain ${\bf b}_{m}$ using \eqref{b_vector}.
This algorithm can be summarized as:
\vspace{-.3cm}
\begin{mdframed}
{\bf Input:} ${\bf H} \in \mathbb{C}^{n\times n}$, and $P$.\\
{\bf Output:} A set of $2n$ candidates for the rows of ${\bf A}$.
\begin{enumerate}
 \item{} Form the matrix ${\bf G}$ as in~(\ref{G}).
 \item{} Reduce ${\bf G}$ to ${\bf G}^{\mbox\small clll}$ by CLLL and compute ${\bf U}={\bf G}^{\mbox\small clll}{\bf G}^{-1}$.
 \item{} Use ${\bf u}_{m}=\left[{\bf d}_{m} |{\bf c}_{m}\right]$, choose ${\bf a}_{m}={\bf c}_{m}$ for $1\leq m\leq 2n$ and put ${\bf b}_{m}$ as in \eqref{b_vector}.
\end{enumerate}
\end{mdframed}
\vspace{.1cm}
\begin{remark}
It is important to highlight the difference between Algorithm $1$ and Algorithm $2$. In Algorithm $1$, a set of candidates for ${\bf a}_{m}$ are chosen via the CLLL reduction technique on a $n$-dimensional complex lattice. %and then ${\bf b}_m$'s are obtained based on the formula given in \eqref{b_vector}.
However, in Algorithm $2$, we jointly obtain ${\bf d}_{m}$ and ${\bf c}_{m}$ on a $2n$-dimensional complex lattice, and only use ${\bf a}_{m} = {\bf c}_{m}$ as candidates for the rows of ${\bf A}$.  Note that vectors ${\bf a}_{m}$ delivered from Algorithm $2$ can be different from that of Algorithm $1$, since
%, the two algorithms attempt to solve minimization of different objective functions
Algorithm $2$ attempts to minimize $g({\bf a}, {\bf b})$, whereas Algorithm $1$ attempts to solve minimization of ${\bf a} {\bf V}{\bf D}{\bf V}^h {\bf a}^{h}$.
\end{remark}
%In general, one can find all rows of ${\bf A}$ via exhaustive search.
Note that, the CLLL algorithm guarantees the existence of at least $n$ vectors of the lattice $\Lambda$ such that the vector in the first row of $\textbf{R}$ is most likely the shortest vector of the lattice.
%\footnote{For $n = 2$, the first row of $\textbf{R}$ is guaranteed to be the shortest vector.}
%and the energy of other vectors are upper bounded by a scalar multiple of
%the successive minima energies. These vectors give us all row vectors
%of ${\bf A}$ and ${\bf B}$ using \eqref{am_clll} and \eqref{b_vector}.
%%%%%%%%%%%%%%%%%%%%%%%%%%%%%%%%%%%%%%%%%%%%%%%%%%%%%%%%%%%%%%%%%%%%%%%%%%%%%%%%%%%%%%
%%%%%%%%%%%%%%%%%%%%%%%%%%%%%%%%%%%%%%%%%%%%%%%%%%%%%%%%%%%%%%%%%%%%%%%%%%%%%%%%%%%%%%
\subsection{Algorithm $3$ via SVD}
For large values of $P$, we can~\cite{zhan12} write ${\bf a}_m{\bf V}{\bf D}{\bf V}^h{\bf a}_m^h$ in \eqref{opt_problem} as $\rho^{-2}_{1}\left\|{\bf v}_{1}{\bf a}_m^h\right\|^2+ \cdots + \rho^{-2}_{n}\left\|{\bf v}_{n}{\bf a}_m^h\right\|^2$, where ${\bf v}_{k}$ denotes the $k$-th row of ${\bf V}$. From the SVD property, we have $\rho_{min} = \rho_{1}$ and $\rho_{max} = \rho_{n}$. If $\rho_{max} \gg \rho_{j}$ for $j \neq n$, then the above equation suggests us to select all ${\bf a}_m$, $1\leq m\leq n$, along the direction of ${\bf v}_{n}$, and as short as possible. On the other hand, if $\rho_{k}$ is large for each $k$, and are comparable, then a set of good candidates can come from choosing a complex integer vector along each ${\bf v}_{k}$.
\vspace{-.3cm}
\begin{mdframed}
{\bf Input:} ${\bf H} \in \mathbb{C}^{n\times n}$, and $P$.\\
{\bf Output:} A set of $n$ candidates for the rows of ${\bf A}$.
\begin{enumerate}
 \item Obtain the SVD of ${\bf H}$ as ${\bf H} = {\bf U\Sigma }{\bf V}^h$.
 \item Choose ${\bf a}_m =\lfloor{\bf v}_m\rceil$ for $1 \leq m \leq n$.
\end{enumerate}
\end{mdframed}
\vspace{.1cm}
 Therefore, another $n$ possible choices for integer vectors can be obtained from the rows of ${\bf V}$ as ${\bf a}_m=\lfloor{\bf v}_m\rceil$ for $1\leq m \leq n$, where $\lfloor \cdot \rceil$ denotes the nearest integer of a real number. %This method can be summarized as:
%%%%%%%%%%%%%%%%%%%%%%%%%%%%%%%%%%%%%%%%%%%%%%%%%%%%%%%%%%%%%%%%%%%%%%%%%%%%%%%%%%%%%%
%%%%%%%%%%%%%%%%%%%%%%%%%%%%%%%%%%%%%%%%%%%%%%%%%%%%%%%%%%%%%%%%%%%%%%%%%%%%%%%%%%%%%%
\subsection{Combined CLLL-SVD solution}
Till now, we have proposed three different low-complexity algorithms to obtain candidate vectors
for the rows of ${\bf A}$. Algorithm $1$ gives us $n$ choices, Algorithm $2$ delivers $2n$ choices, whereas Algorithm $3$ brings another $n$ possible choices. Overall, we can use all the $4n$ candidate vectors for ${\bf a}_m$ and obtain the corresponding vectors
${\bf b}_m$, $1\leq m\leq 4n$ using the relation in \eqref{b_vector}. With this, we have $4n$ pairs of $({\bf a}_m,{\bf b}_m)$. Now, we proceed to sort these $4n$ vectors in the increasing order of their
$g({\bf a}_m, {\bf b}_m)$ values. Then, we find the first $n$ vectors %the best\footnote{The word ``best'' means smallest in terms of $g(.,.)$ function}
which form an invertible matrix over the ring $\mathcal{R}$. Note that all the three algorithms have low computational complexity. As a result, the combining algorithm has lower computational complexity than the exhaustive search and other lattice reduction algorithms like HKZ and Minkowski, and hence, the proposed technique is amenable to implementation. We refer to the combined solution as ``combined CLLL-SVD solution".

\subsection{Complexity Comparison}\label{sec:comparison}
It is clear that the combined CLLL-SVD method includes at most a SVD algorithm and a matrix inversion with complexity $O\left(n^3\right)$, a $2n$-dimensional CLLL algorithm with complexity $O\left(4n^2\log(2n)\right)$ following by a sorting algorithm of size $4n$ with complexity $O\left(4n\log(4n)\right)$. Therefore the complexity of this approach is $O\left(n^3\right)$ and it is obviously independent of $P$. The complexity of the proposed exhaustive search in~\cite{zhang12} is known to be $O\left(P^n\right)$. The complexity of other reduction algorithms like HKZ and Minkowski are given in~\cite{SHV:submitted}. In Table~\ref{table:complexitycompare}, we compare the complexity of finding matrix ${\bf A}$ for different known methods.
\begin{table*}
\caption{\label{table:complexitycompare} Complexity Comparison.}
\begin{center}
\begin{tabular}{||c|c|c||} \hline\hline
\multirow{1}{*}{Name} & Approach & Complexity\\
\hline\hline
\multirow{4}{*}{Integer-Forcing Linear receiver} & Brute Force & $P^n$ \\
& HKZ reduction & $(2\pi e)^{n+O(\log 2n)}$\\
& Minkowski reduction & $(5/4)^{2n^2}$\\
& Combined CLLL-SVD solution & $n^3$\\
\hline\hline
\end{tabular}
\end{center}
\end{table*}

\section{Simulation Results}~\label{sec:simulations}
\label{sec_4}
\begin{figure}[t]%
  \begin{center}%
\includegraphics[width=7.6cm]{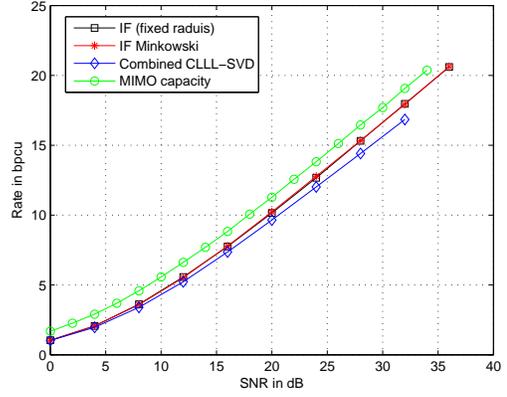}~\caption{\label{fig:ErgodicRateCapacity22}
Ergodic rate of various linear receivers for $2 \times 2$ MIMO channel.}
  \end{center}
\end{figure}
%\vspace{-.1cm}
In this section, we present simulation results on the ergodic rate (see~\cite{SHV:submitted} or~\cite{sakzad12} for definition of ergodic rate) and the probability of error for the following receiver architectures on $2 \times 2$ MIMO channel: ({\em i}) IF linear receiver with exhaustive search, ({\em ii}) IF linear receiver with Minkowski lattice reduction solution~\cite{SHV:submitted}, ({\em iii}) IF linear receiver with combined CLLL-SVD solution, ({\em iv}) the ZF and MMSE linear receiver, and ({\em v}) the joint maximum likelihood (ML) decoder. For the IF receiver with exhaustive search, the results are presented with the constraint of fixed radius for the exhaustive search. We have not used the radius constraint given in \eqref{radius} as the corresponding search space increases with $P$. Instead, we have used a fixed radius of $8$ for all values of $P$. By relaxing this constraint, we have reduced the complexity of brute force search, noticeably.

In Fig. \ref{fig:ErgodicRateCapacity22}, we present the ergodic rate of the above listed receivers, wherein, for the case of ML receiver, the ergodic capacity~\cite{Telatar99} of $2 \times 2$ MIMO channel has been presented. The rate for IF linear receiver with Minkowski lattice reduction solution is cited from~\cite{SHV:submitted}. We observe that the combined CLLL-SVD solution performs pretty much the same as IF receiver with exhaustive search and IF based on Minkowski lattice reduction solution at low and moderate $\mbox{\small SNR}$s. %In low ${\small \mbox{SNR}}$s, MMSE receiver performs better than IF while at high ${\small \mbox{SNR}}$s IF receiver with exhaustive search beats MMSE receiver.
Also, note that the combined CLLL-SVD and Minkowski lattice reduction solutions give lower bounds on the ergodic rate of exhaustive search based IF receiver while the latter one is tighter. For the ergodic rate results of the IF receiver, we have used ${\bf A}$ matrices which are invertible over $\mathbb{Z}[i]$.

\indent Now, we present the uncoded bit error rate (BER) for the above receiver architectures with $4$-QAM constellation.
%At the transmitter side, a complex symbol $x_{j} \in \mathcal{S'}$ (where $\mathcal{S'}$ is a finite
%constellation) is transmitted through the $j$-th transmit antenna for each $j = 1, 2$.
We use the finite constellation $\mathcal{S} = \{ 0, 1, i, 1+i \}$ carved out of the infinite lattice $\mathbb{Z}[i]$. Note that $\mathbb{Z}[i] = \mathcal{S} \oplus 2\mathbb{Z}[i]$. In this method, the received vector is of the form, ${\bf y} = \sqrt{\frac{\mbox{\small SNR}}{4}}{\bf H}{\bf s} + {\bf z},$
where ${\bf s} \in \mathcal{S}^{2 \times 1}$. For the above setting, we use modulo lattice decoding at the receiver. Each component of ${\bf y}$ is decoded to the nearest point in $\mathbb{Z}[i]$ and then ``modulo $2$" operation is performed independently on its in-phase and quadrature component. With this, we get ${\bf r} \in \mathcal{S}^{2 \times 1}$ from both the components of ${\bf y}$. Further, we solve the system of linear equations ${\bf r} = {\bf A}{\bf s} \mbox{ modulo 2}$ over the ring $\mathcal{R}$ to obtain the decoded vector $\hat{{\bf s}}$.
%\end{itemize}
\begin{figure}[t]%
  \begin{center}%
  %\vspace{1.5cm}
  %\label{prob_4qam}
\includegraphics[width=7.6cm]{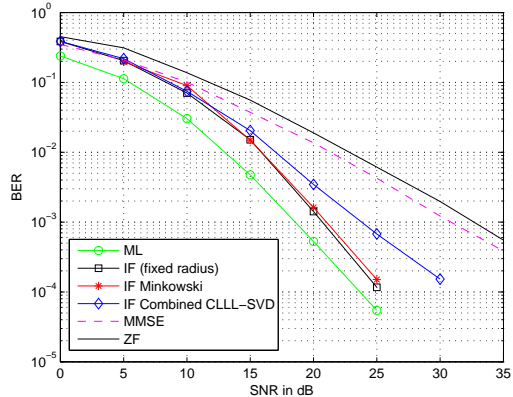}~\caption{\label{fig:QPSK22}BER for various linear receivers with $4$-QAM constellation.}
  \end{center}
\end{figure}
%\vspace{-.1cm}
In Fig. \ref{fig:QPSK22}, we present the BER results for all the five receiver architectures. Note that both the IF receiver with combined CLLL-SVD and Minkowski lattice reduction solutions outperform the ZF and MMSE architectures, but trades-off error performance for complexity in comparison with brute force search. In particular, the combined CLLL-SVD approach fails to provide diversity results as that of the exhaustive search and Minkowski lattice reduction approaches. This diversity loss is due to the larger value of $\max_{m}g({\bf a}_m,{\bf b}_m)$ delivered by the combined CLLL-SVD algorithm in comparison with the optimum solution. For the probability of error results of the IF receiver, we have used ${\bf A}$ matrices which are invertible over $\mathbb{Z}_{2}[i]$. We have observed similar results for $4\times4$ MIMO channels as well.

\section{Conclusions}~\label{sec:conclusion}%${
In~\cite{SHV:submitted}, Algorithm $1$ along with HKZ and Minkowski lattice reduction algorithms was employed to find the matrix ${\bf A}$. Algorithm $1$ which includes the complex LLL method as an alternate technique for the above reduction methods turned out to be not satisfactory in terms of ergodic rate and error performance. Hence, we have tried to improve effectiveness of CLLL algorithm by combining it with other approaches such as Algorithms $2$ and $3$. We have proposed a low-complexity systematic method called the combined CLLL-SVD algorithm for the MIMO IF architecture. Simulation results on the ergodic rate and the probability of error were also presented to reveal the effectiveness of combined CLLL-SVD solution versus other linear receivers. The proposed combined algorithm trades-off error performance for complexity in comparison with both IF receivers based on exhaustive search and Minkowski or HKZ lattice reduction algorithms. Further improvements are required to achieve results which are competitive with IF receivers based on exhaustive search and the ones presented in~\cite{SHV:submitted}.
%In future work, we are exploring for other low-complexity solutions, which can provide full diversity property.
%\section{Acknowledgments}
%$


\begin{thebibliography}{1}
%\bibitem{Avestimehr11} S.~Avestimehr, S.~Diggavi, and D.~Tse,
%``Wireless network information flow: A deterministic approach,''
%{\em IEEE Trans. Inf. Theory,} vol.~57, pp.~1872--1905, 2011.

%\bibitem{Biglieri98} E.~Biglieri, J.~Proakis, and S.~Shamai,
%``Fading channels: information-theoretic and communications aspects,"
%{\em IEEE Trans. Inf. Theory}, vol.~44, No.~6, pp.~2619--2692, 1998.

\bibitem{Feng} C.~Feng, D.~Silva, and F.R.~Kschichang,
``An algebraic approach to physical-layer network coding,''
{\em submitted to IEEE Trans. Inform. Theory}, arXiv: 1108.1695v1.

\bibitem{CLLL09} Y.H.~Gan, C.~Ling, and W.H.~Mow,
``Complex lattice reduction algorithm for low-complexity full-diversity MIMO detection,''
{\em IEEE Trans. Signal Processing}, vol.~57, pp.~2701--2710, 2009.

%\bibitem{Lagarias90}
%J.~Lagarias, H.~Lenstra Jr., and C.~Schnorr,
%``Korkin-Zolotarev bases and successive minima of a lattice and its reciprocal lattice,''
%{\em Combinatorica}, vol.~10, no.~4, pp.~333--348, 1990.

%\bibitem{Liew11} S.-C.~Liew, S.~Zhang, and L.~Lu,
%``Physical-layer network coding: Tutorial, survey, and beyond,''
%{\em Phys. Commun., 2011} [Online]. Available: http://arxiv.org/abs/1105.4261.

%\bibitem{Lim11} S.H.~Lim, Y.-H.~Kim, A.~El Gamal, and S.-Y.~Chung,
%``Noisy network coding,''
%{\em IEEE Trans. Inf. Theory}, vol.~57, pp.~3132--3152, 2011.

%\bibitem{Narayanan07}
%K.~Narayanan, M.P.~Wilson, and A.~Sprintson, ``Joint physical layer
%coding and network coding for bi-directional relaying," in Proc.~45th
%{\em Annual Allerton Conference on Communications, Control and Computing}, Monticello, IL, Sep.~2007.

\bibitem{Nazer11} B.~Nazer and M.~Gastpar,
``Compute-and-Forward: harnessing interference through structured codes,''
{\em IEEE Trans. Inform. Theory}, vol.~57, pp.~6463--6486, 2011.

%\bibitem{Niesen} U.~Niesen, B.~Nazer, and P.~Whiting,
%``Computation Alignment: Capacity Approximation without Noise Accumulation,"
%{\em submitted to IEEE Transactions on Information Theory}, arXiv: 1108.6312v1.

\bibitem{Narayanan10}
M.P.~Wilson, K.~Narayanan, H.~Pfister, and A.~Sprintson, ``Joint physical
layer coding and network coding for bidirectional relaying,''
{\em IEEE Trans. Inform. Theory}, vol.~11, pp.~5641--5654, 2010.

%\bibitem{belfiore} A.~Osmane and J.C.~Belfiore,
%``The Compute-and-Forward Protocol: Implementation and Practical Aspects,"
%{\em submitted to IEEE Communications Letters}, arXiv:1107.0300v1.

%\bibitem{TL} A.~Sakzad, M-R.~Sadeghi, and D.~Panario,
%``Turbo Lattices: Construction and Performance Analysis",
%{\em submitted to IEEE Transactions on Information Theory}, arXiv: 1108.1873v1.

\bibitem{korkine1873} A.~Korkine, and G.~Zolotareff,
``Sur les formes quadratiques,''
{\em Math. Ann.}, vol.~6, pp. 366--389, 1873.

\bibitem{Kumar09}
K.~Kumar, G.~Caire, and A.~Moustakas, ``Asymptotic performance
of linear receivers in MIMO fading channels,''
{\em IEEE Trans. Inform. Theory}, vol.~55, pp.~4398--4418, 2009.

\bibitem{Minkowski1891} H.~Minkowski,
``\"{U}ber die positiven quadratischen formen und \"{u}ber kettenbruch\"{a}hnliche algorithmen,''
{\em J. Reine und Angewandte Math.}, vol.~107, pp. 278--297, 1891.

\bibitem{SHV:submitted} A.~Sakzad, J.~Harshan, and E.~Viterbo,
``Integer-forcing linear receivers based on lattice reduction algorithms,''
{\em submitted to IEEE Trans. Wireless Communications}, arXiv:1209.6412.

\bibitem{sakzad12} A.~Sakzad, E.~Viterbo, Y.~Hong, and and J.~Boutros,
``On the ergodic rate for compute-and-Forward,"
{\em Proceeding of International Symposium on Network Coding 2012 (NetCod 2012), MIT University, Boston, MA, USA}.

\bibitem{Taherzadeh07-1} M.~Taherzadeh, A.~Mobasher, and A.~Khandani,
``Communication over MIMO broadcast channels using lattice-basis reduction,''
{\em IEEE Trans. on Inform. Theory,} vol.~53, pp.~4567--4582, 2007.

\bibitem{Taherzadeh07-2} M.~Taherzadeh, A.~Mobasher, and A.~Khandani,
``LLL reduction achieves the receive diversity in MIMO decoding,''
{\em IEEE Trans. on Inform. Theory,} vol.~53, pp. 4801--4805, 2007.

\bibitem{Telatar99}
I.~Telatar, ``Capacity of multi-antenna Gaussian channels,''
{\em European Trans. Telecommun.}, vol.~10, pp.~586--595, 1999.

%\bibitem{viterbo11} E.~Viterbo, Y.~Hong, and and J.~Boutros,
%``Wireless network coding over finite rings,"
%{\em Workshop on Algebraic Structure in Network Information Theory},
%Banff International Research Station, Aug.~2011.

\bibitem{matz11} D.~Wubben, D.~Seethaler, J.~Jalden, and G.~Matz,
``Lattice reduction: a survey with applications in wireless communications,''
{\em IEEE Signal Process. Mag.}, vol.~28, pp.~70--91,~2011.

\bibitem{zhan12} J.~Zhan, B.~Nazer, U.~Erez, and M.~Gastpar,
``Integer-forcing linear receivers,''
{\em Information Theory Proceedings (ISIT), 2010 IEEE International Symposium on},
pp.~1022--1026, 2010. Extended version is available at: http://arxiv.org/abs/1003.5966.

\bibitem{zhang12} W.~Zhang, S.~Qiao, and Y.~Wei,
``HKZ and Minkowski Reduction Algorithms for Lattice-reduction-aided MIMO Detection,''
{\em to appear in IEEE Trans. on Signal Processing}.
Available at: http://ieeexplore.ieee.org/stamp/stamp.jsp?tp=$\&$arnumber=6256756.

\end{thebibliography}
\end{document}